\documentclass[epsf,psfig]{mn2e}

\usepackage{graphicx}

\newcommand{\ms}{\mbox{m s$^{-1}~$}}

\newcommand{\msun}{M$_{\odot}$}

\newcommand{\mjup}{$M_{\rm Jup}$}

\newcommand{\msini}{$M \sin i~$}

\newcommand{\logrhk}{log $R'_{\rm HK}$}

\begin{document}

\title[Anglo-Australian Planet Search]
{High eccentricity planets from the Anglo-Australian Planet Search}

\author[H. R. A. Jones et al.]{
\parbox[t]{\textwidth}{
Hugh R. A. Jones$^1$, R. Paul Butler$^2$,  
C.G. Tinney$^3$, Geoffrey W. Marcy$^4$,  
Brad D. Carter$^5$, Alan J. Penny$^{6,7}$,
Chris McCarthy$^8$, Jeremy Bailey$^9$} \\\\
%\vspace*{6pt} 
$^1$Centre for Astrophysics Research, University of Hertfordshire, 
College Lane, Hatfield, Herts AL10 9AB, UK\\
$^2$Carnegie Institution of Washington, Department of 
Terrestrial Magnetism, 5241 Broad Branch Rd NW, Washington, 
DC 20015-1305, USA \\
$^3$Anglo-Australian Observatory, PO Box 296, Epping. 1710, Australia\\  
$^4$Department of Astronomy, University of California, Berkeley, CA, 94720, USA\\
$^5$Faculty of Sciences,  University of Southern Queensland, Toowoomba, 
QLD 4350, Australia\\
$^6$Rutherford Appleton Laboratory, Chilton, Didcot, Oxon OX11 0QX, UK\\
$^7$SETI Institute, 515 N. Whisman Road, Mountain View, CA 94043, USA\\
$^8$Department of Physics and Astronomy, San Francisco State University, 
San Francisco, CA 94132, USA\\
$^9$Australian Centre for Astrobiology, Macquarie University, Sydney NSW 2109, Australia\\}

\date{}

\maketitle

%\label{firstpage}

\begin{abstract}
We report Doppler measurements of the stars HD~187085 and HD~20782 which 
indicate two high eccentricity low-mass companions to the stars.  
We find HD~187085 has a Jupiter-mass companion with a $\sim$1000~d orbit.
Our formal `best fit' solution suggests an eccentricity of 0.47, however, 
it does not sample the periastron passage of the companion and we find
that orbital solutions with eccentricities between 0.1 and 0.8  give 
only slightly poorer fits (based on RMS and $\chi_{\nu}^2$) 
and are thus plausible. 
Observations made during periastron passage in 
2007 June should allow for the reliable determination of the orbital eccentricity 
for the companion to HD~187085.
Our dataset for HD~20782 does sample periastron
and so the orbit for its companion can be more reliably determined. We find 
the companion to HD~20782 has $M$~sin~$i$~=~1.77$\pm$0.22 \mjup, an orbital period of 595.86$\pm$0.03~d and an orbit with an
eccentricity of 0.92$\pm$0.03. 
%We note that this high eccentricity solution depends on 
%observations at one epoch without which a somewhat 
%lower eccentricity would be fit. 
The detection of such high-eccentricity (and relatively low velocity 
amplitude)
exoplanets appears to be facilitated by the long-term precision of 
the Anglo-Australian Planet Search.
Looking at exoplanet detections as a whole, 
we find that those with higher eccentricity 
seem to have relatively higher velocity amplitudes 
indicating higher mass planets and/or an observational bias against the 
detection of high eccentricity systems.
\end{abstract}

\begin{keywords}
planetary systems - stars: individual (HD187085) (HD20782), 
brown dwarfs
\end{keywords}

\section{Introduction}
The Anglo-Australian Planet Search (AAPS) is a long-term radial velocity
project engaged in the detection and measurement of 
extra-solar planets (hereafter, shortened to exoplanets) 
at the highest possible precisions.
Together with programmes using similar techniques on the
Lick 3\,m and Keck I 10\,m telescopes (Fischer et al. 2001; Vogt et al. 2000),
it provides all-sky planet search coverage for inactive F, G, K and M dwarfs
down to a magnitude limit of V=7.5. So far the AAPS has  
has published data for 23 exoplanets (Butler et al. 2001; 
Carter et al. 2003; Jones et al. 2002a,b, 2003; McCarthy et al. 2005;
Tinney et al. 2001, 2002a, 2003, 2004, 2005, 2006) as well as non-detections
of planets claimed elsewhere (Butler et al. 2001; Butler et al. 2002a).

The AAPS is carried out on the 3.9m
Anglo-Australian Telescope (AAT) using the University College London Echelle
Spectrograph (UCLES), operated in its 31 lines/mm mode
together with an I$_{2}$ absorption cell. 
UCLES now uses the AAO's EEV 2048$\times$4096 13.5$\mu$m pixel CCD,
which provides excellent quantum efficiency across the 500--620~nm
I$_2$ absorption line region. From 1998 until 2002, the AAPS 
allocation was around 20 nights per year, from 2002 to 2004 it increased
to 32. In 2005 the AAPS has an allocation of 64 nights per year. 
Despite this search taking place on a common-user telescope
with frequent changes of instrument, we have historically
achieved a 3~m~s$^{-1}$ precision
down to the V~=~7.5 magnitude limit of the survey (e.g.,
fig. 1, Jones et al. 2002a), for stars with low `jitter'. In addition
to our published errors, there are additional sources of error 
collectively termed `jitter'. The magnitude of the `jitter' is a 
function  of the spectral type of the star observed. Wright (2005) gives a model
which estimates the jitter for a star based upon its activity.
Our continuing attainment of 3~m~s$^{-1}$ single-shot precision for
suitably inactive stars is demonstrated in McCarthy et al. (2004) and  
Tinney et al. (2005).

Our target sample, which we have observed since 1998, is given in
Jones et al. (2002b). It includes 178 late (IV-V) F, G and K
stars with declinations below $\sim -20^\circ$ and is complete
to V$<$7.5. We also observe sub-samples of 20 metal-rich
([Fe/H]$>$0.3) stars with V$<$9.5 and 7 M dwarfs with V$<$11 and
declinations below $\sim -20^\circ$. The sample has been increased
to more than 250 solar-type stars to be complete to a magnitude limit of V=8.
Where age/activity information is
available from \logrhk indices (Henry et al. 1996; Tinney et al. 2002b;
Jenkins et al. 2006)
we require target stars to have log $R'_{\rm HK}$ $<$ --4.5
corresponding to
ages typically greater than 3 Gyr. Stars with known stellar companions within
2 arcsec are removed from the observing list, as it is operationally difficult
to get an uncontaminated spectrum of a star with a nearby companion.
Spectroscopic binaries discovered during the programme have also been
removed and are reported elsewhere (Blundell et al. 2006).
Otherwise there is no bias against observing multiple stars. The programme
is also not expected to have any bias against brown dwarf companions.

The observing and data processing procedures follow those described by 
Butler et al. (1996, 2001) and Tinney et al. (2005).  
All data taken by the AAPS to date have been
reprocessed through our continuously upgraded anlaysis system. Here, 
we report results from the current version of our pipeline.

\begin{figure}
\includegraphics[width=55mm,angle=-90]{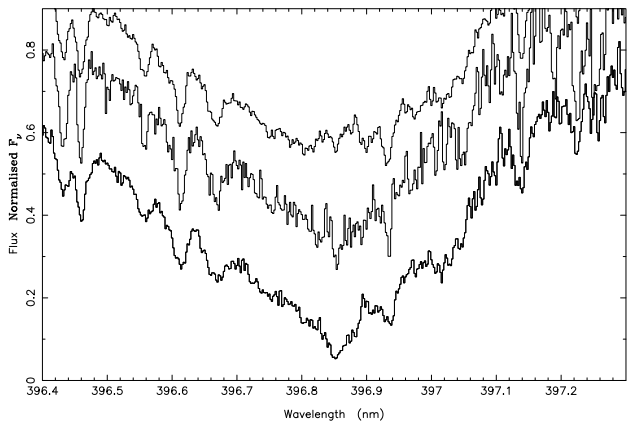}
\caption{UCLES spectrum of the region of the CaHK line for 
HD~20782 (middle, log $R$'$_{\rm HK}$ = --4.93, G3V) 
and two comparison objects of similar spectral types and activities
HD~17051 (top, log $R$'$_{\rm HK}$ = --4.81, G0V)
and HD~142 (bottom, log $R$'$_{\rm HK}$ = --4.95, G0V). The spectra
are taken from Tinney et al. (2002).
}
\label{cahk}
\end{figure}

\begin{table}
 \centering
% \begin{minipage}{140mm}
\caption{The stellar parameters for HD~187085 and HD~20782 tabulated below
are taken from a range of sources: spectral 
types from Houck (1982), activities from Henry et al. (1996) and Tinney et al.
(2002); Hipparcos measurements from (ESA 1997); luminosities, masses, temperatures,
metallicities and space velocities from Nordstrom et al. (2004, abbreviated
as N04 in the table, the errors given for Nordstrom et al. values are the 
average of dispersions, no upper confidence limit is given for the age of 
HD~20782 by Nordstrom et al.) and Fischer \& Valenti (2005, abbreviated
as FV05 in the table); jitter values are taken from 
Wright (2005).}
\label{stellarp}
  \begin{tabular}{@{}lrrr@{}}
Parameter & HD~187085 & HD~20782 \\
& & \\
Spectral Type & G0V &  G3V \\
log $R'_{\rm HK}$ & -4.93 & -4.91 \\
Hipparcos $N_{\rm obs}$ & 113 & 191\\
Hipparcos $\sigma$ & 0.0013 & 0.0009\\
log (L$_{\rm star}$/L$_\odot$) & 2.13$\pm^{0.55}_{0.44}$ &
1.25$\pm^{0.2}_{0.17}$ \\
M$_{\rm star}$/M$_\odot$ -- N04& 1.16$\pm^{0.06}_{0.08}$ & 0.90$\pm^{0.07}_{0.03}$ \\
M$_{\rm star}$/M$_\odot$ -- VF05& 1.22$\pm^{0.1}_{0.06}$ & 1.0$\pm$0.03 \\
T$_{\rm eff}$ (K)-- N04   &    6011$\pm$76   &         5636$\pm$76 \\
T$_{\rm eff}$ (K) -- VF05     &    6075$\pm$30   &         5758$\pm$30\\
$[$Fe/H$]$ -- N04 & 0.05$\pm$0.10 & -0.16$\pm$0.10 \\
$[$Fe/H$]$ -- VF05   & 0.05$\pm$0.03     &   -0.05$\pm$0.03\\
U,V,W -- N04 (km~s$^{-1}$) &11,-20,-14 & -38,-61,-2 \\
Age (Gyr) -- N04 & 3.9$\pm^{2.2}_{1.4}$ & 13.0\_$_{4.5}$\\
Age (Gyr) -- VF05 & 3.3$\pm^{0.6}_{1.2}$ & 7.1$\pm^{1.9}_{4.3}$ \\
jitter  (m~s$^{-1}$)  &  2.80              &        4.11\\
%$M_{\rm V}$ & 3.95$\pm$xx& 4.57$\pm$xx\\
%$M_{\rm bol}$ & & \\
%R$_{\rm star}$    &  1.32$\pm$0.05 &       1.13$\pm$0.05\\
%Log g    & 4.28$\pm$0.08   &     4.35$\pm$0.08\\
%V~sin~i    &  5.09$\pm$0.5    &      2.36$\pm$0.5\\
%distance         &   45.0$\pm$2.3           &         36.0$\pm$1.1     \\
%V & -20& -61 \\
%W & -14& -2 \\
%Hipparcos $\pi$ & 22.2$\pm$1.1 & 27.8$\pm$0.9\\
\end{tabular}
%\end{minipage} 
\end{table}

\begin{table}
 \centering
% \begin{minipage}{140mm}
\caption{Relative radial velocities are given for HD~187085. Julian
Dates (JD) are heliocentric.
RVs are barycentric but have an arbitrary
zero-point determined by the radial velocity of the template.}
\label{vel187085}
  \begin{tabular}{@{}lrc@{}}
JD & RV & error \\
(-2451000)   &  (m s$^{-1}$) & (m s$^{-1}$)\\
& & \\
   120.9170  &   -12.1  &  5.3 \\
   411.0753  &    -2.5  &  6.5 \\
   683.1693  &    16.1  &  5.7 \\
   743.0494  &     9.3  &  5.5 \\
   767.0046  &     8.8  &  4.7 \\
   769.0652  &     5.3  &  4.4 \\
   770.1153  &     5.3  &  5.2 \\
   855.9477  &     8.6  &  8.9 \\
  1061.2140  &    -4.5  &  5.3 \\
  1092.0512  &   -28.2  &  5.9 \\
  1128.0267  &   -25.9  &  5.1 \\
  1151.0146  &   -15.4  &  6.8 \\
  1189.9230  &    -6.4  &  4.2 \\
  1360.2816  &    -7.3  &  4.6 \\
  1387.2168  &    -8.7  &  3.5 \\
  1388.2355  &   -12.1  &  3.7 \\
  1389.2664  &   -12.9  &  4.3 \\
  1422.2121  &    -6.9  &  4.0 \\
  1456.0917  &    -7.9  &  4.9 \\
  1750.2587  &    13.5  &  3.8 \\
  1752.2311  &    11.9  &  3.8 \\
  1784.2071  &    19.6  &  3.2 \\
  1857.1218  &    -4.2  &  3.7 \\
  1861.0168  &     9.2  &  4.5 \\
  1942.9842  &    17.2  &  4.2 \\
  1946.9260  &    17.9  &  3.2 \\
  2217.0666  &   -12.1  &  3.7 \\
  2245.0429  &   -24.0  &  5.9 \\
  2484.3002  &    -3.0  &  3.7 \\
  2489.2600  &    -5.2  &  3.4 \\
  2507.1966  &    -0.7  &  2.8 \\
  2510.2154  &    -2.1  &  3.3 \\
  2517.2619  &    -3.5  &  4.1 \\
  2520.2747  &   -10.4  &  4.2 \\
  2569.0854  &    -5.8  &  3.9 \\
  2572.1667  &    -8.3  &  4.5 \\
  2577.0269  &     2.1  &  3.0 \\
  2627.9629  &     9.8  &  5.8 \\
  2632.0744  &     6.7  &  3.4 \\
  2665.9534  &    18.9  &  3.7 \\
\end{tabular}
%\end{minipage}
\end{table}

\begin{table}
 \centering
% \begin{minipage}{140mm}
\caption{Relative radial velocities (RV) are given for HD~20782. Julian
Dates (JD) are heliocentric.
RVs are barycentric but have an arbitrary
zero-point determined by the radial velocity of the template.}
\label{vel20782}
  \begin{tabular}{@{}lrc@{}}
JD & RV & error \\
(-2451000)   &  (m s$^{-1}$) & (m s$^{-1}$) \\
& & \\
    35.3195  &    24.3  &  4.6 \\
   236.9306  &     4.8  &  6.0 \\
   527.0173  &     9.2  &  6.8 \\
   630.8824  &    31.7  &  5.3 \\
   768.3089  &    -3.9  &  5.1 \\
   828.1107  &    -4.8  &  6.0 \\
   829.2745  &    -0.5  &  8.3 \\
   856.1353  &    -2.9  &  7.5 \\
   919.0066  &     6.5  &  6.1 \\
   919.9963  &     6.3  &  5.7 \\
   983.8901  &    11.8  &  6.6 \\
  1092.3044  &    20.3  &  4.7 \\
  1127.2681  &    19.6  &  5.6 \\
  1152.1631  &    21.4  &  5.1 \\
  1187.1597  &    25.5  &  4.9 \\
  1511.2066  &    -0.3  &  4.5 \\
  1592.0484  &    19.1  &  4.5 \\
  1654.9603  &    22.3  &  4.4 \\
  1859.3054  &  -197.6  &  3.8 \\
  1946.1383  &   -14.9  &  4.0 \\
  1947.1225  &   -12.3  &  3.4 \\
  2004.0015  &     3.1  &  3.6 \\
  2044.0237  &     7.6  &  4.3 \\
  2045.9607  &     4.8  &  3.8 \\
  2217.2881  &    12.1  &  3.3 \\
  2282.2204  &    26.5  &  3.8 \\
  2398.9697  &    27.8  &  2.6 \\
  2403.9607  &    30.3  &  4.8 \\
  2576.3073  &    -5.9  &  3.0 \\
  2632.2813  &    -3.8  &  3.2 \\
\end{tabular}
%\end{minipage}
\end{table}

\section{Stellar Characteristics of HD~187085 and HD~20782}
The characteristics for HD~187085 and HD~20782 are summarised in Table
\ref{stellarp}. Houck (1982) assigns spectral types of G0V and G3V to
HD~187085 and HD~20782 which are consistent with information derived for
and from the Hipparcos mission (ESA 1997). Both stars have been included
in large-scale studies of nearby solar type stars. Nordstrom et al. (2004)
included them in a magnitude-limited, kinematically unbiased study of
16682 nearby F and G dwarf stars. Valenti \& Fischer (2005) included them 
in a study of the stellar properties for 1040 F, G and K stars observed
for the Anglo-Australian, Lick and Keck planet search programmes. 
Valenti \& Fischer used high signal-to-noise
echelle spectra originally taken for template radial velocity 
spectra and spectral synthesis to derive effective temperatures,
surface gravities and metallicities whereas Nordstrom et al. (2004)
used Str\"{o}mgren photometry and the infrared flux method calibration of 
Alonso et al. (1996). Both studies use Hipparcos parallaxes to convert 
luminosities in order to make comparisons with different theoretical isochrones
to derive stellar parameters. To determine, stellar masses and ages 
Valenti \& Fischer, use Yonsei-Yale isochrones (Demarque et al. 2004) and 
Nordstrom et al. (2004) use Padova isochrones (Giradi et al. 2000; 
Salasnich et al. 2000). Both sets of derived parameters agree to within
errors. 

Both HD~187085 and HD~20782 have moderate activity indices
\logrhk~$\sim$~--4.9 from Henry et al. (1996). For the case of 
HD~187085 we also have a confirmation measurement from our 
own CaHK programme (Fig. \ref{cahk} and Tinney et al. 2002)
Furthermore there is no evidence for significant photometric
variability for either star in measurements made by the Hipparcos satellite.
Combining Hipparcos astrometry with their radial velocities, Nordstrom
et al. (2004) determine U, V, W space velocities. Based on 
Edvardsson (1993) these ages are consistent with the derived ages
in Table \ref{stellarp}.

\begin{table}
 \centering
% \begin{minipage}{140mm}
  \caption{Orbital parameters for the companions to HD~187085 and HD~20782.}
  \begin{tabular}{@{}lrrr@{}}
\label{orbit}
Parameter & HD~187085 & HD~20782 \\
& & \\
Orbital period $P$ (d) &  986 & 585.86$\pm$0.03 \\
Velocity amplitude $K$ (m~s$^{-1}$) & 17 & 115$\pm$12 \\
Eccentricity $e$  & 0.47 & 0.92$\pm$0.03 \\
$\omega$ (deg)    & 94     &   147$\pm$3 \\
Periastron Time (JD) & 2450912 & 2451687.1$\pm$2.5 \\
$M$sin$i$ (\mjup)     &  0.75        & 1.80$\pm$0.23 \\
a (au)            &  2.05        & 1.36$\pm$0.12  \\
RMS (\ms)         &  7.1       & 5.0  \\
%$\chi_{\nu}^2$   &  1.82        & 1.01  \\
\end{tabular}
%\end{minipage}
\end{table}

\begin{figure}
\hspace*{-1cm}
\includegraphics[width=70mm,angle=90]{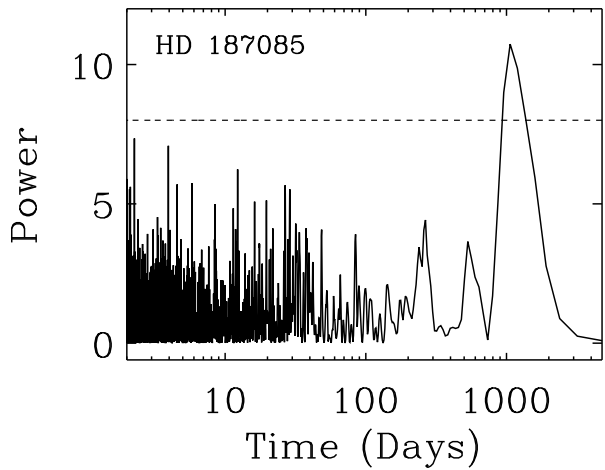}
\caption{A periodogram of the 40 epochs of HD~187085 obtained 
at the AAT between 1998 November and 2005 October.  The false alarm 
probabilities shown are based on the assumption of normally distributed errors.
The dashed line is the 1\% false alarm level.
}
\label{per187085}
\end{figure}

\begin{figure}
%\hspace*{-1cm}
\includegraphics[width=65mm,angle=90]{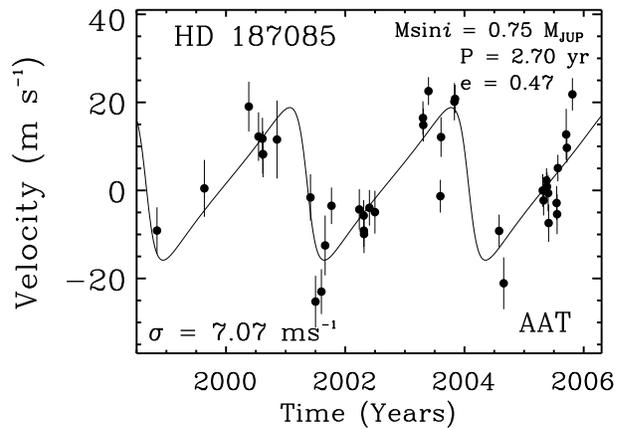}
\includegraphics[width=65mm,angle=90]{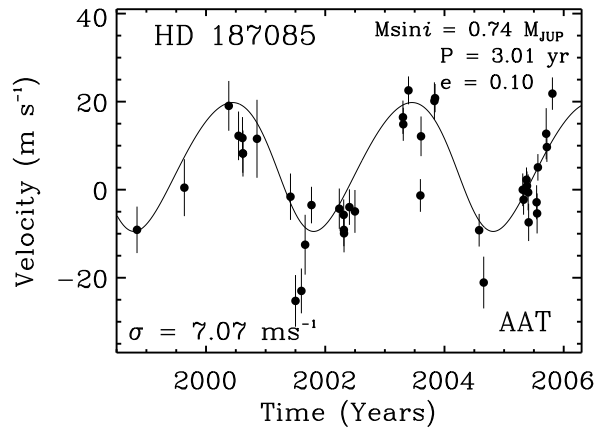}
 \caption{Doppler velocities obtained for HD~187085 from 1998 August to
2005 October. The top plot shows the best fit eccentricity, the bottom plot 
shows a similar quality fit for a fixed eccentricity of 0.1. In both cases
the solid lines indicate the best fit Keplerian with the parameters shown.}
\label{hd187085}
\end{figure}

\begin{figure}
%\vspace{-3cm}
\hspace*{-1cm}
\includegraphics[width=60mm,angle=90]{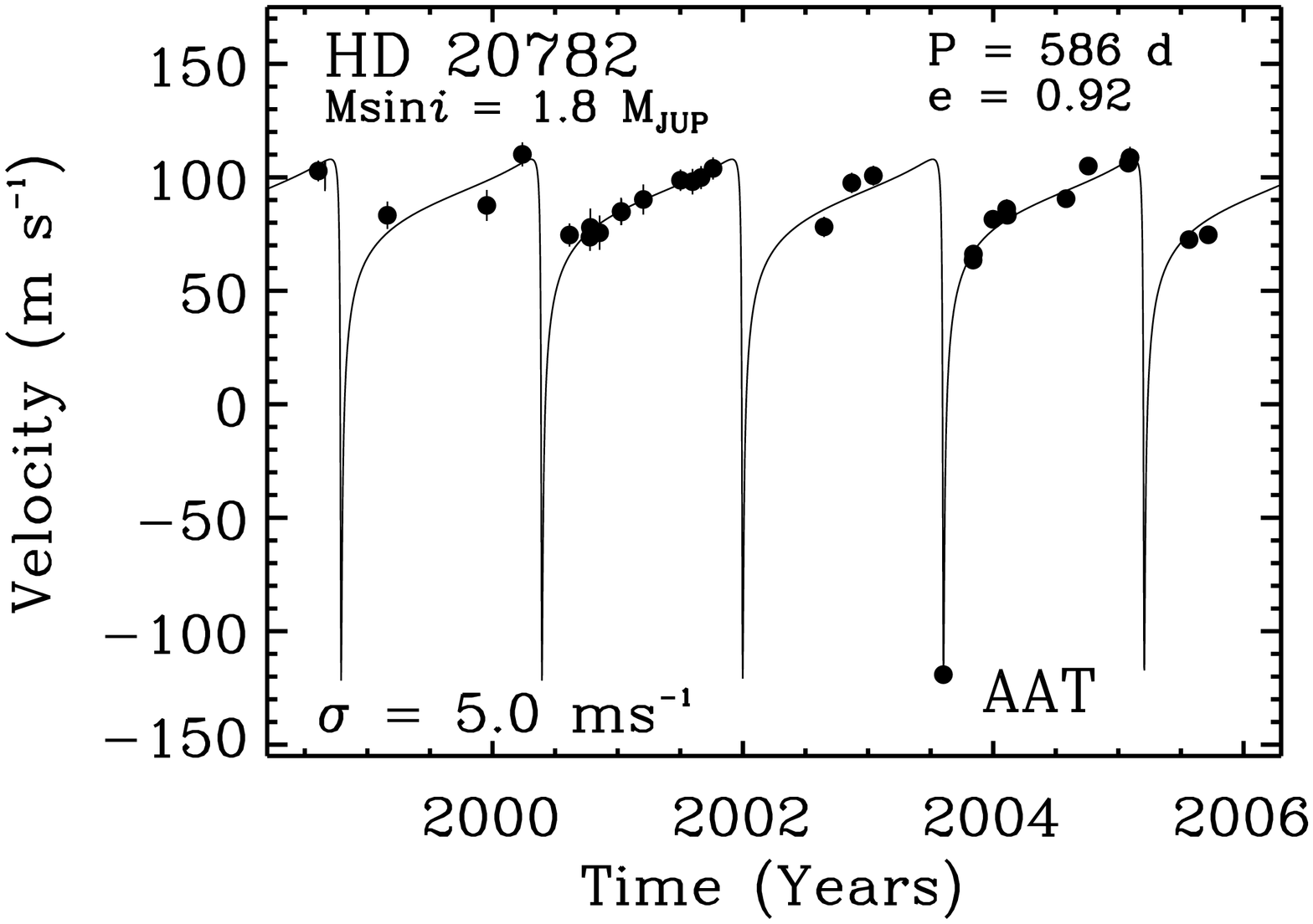}
 \caption{Doppler velocities obtained for HD~20782 from 1998 August to
2005 September.
The solid line is a best fit Keplerian with the parameters shown
in Table \ref{orbit}.
The RMS of the velocities about the fit is 5.0~m~s$^{\rm -1}$.
Based on a primary mass of 0.95~\msun,
the minimum (M~sin~$i$) mass of the companion is 1.77~M$_{\rm JUP}$ and
the semimajor axis is 1.36~au.}
\label{hd20782}
\end{figure}

\section{Orbital Solution for HD~187085}

The 40 epochs of Doppler velocity measurements for HD~187085, obtained
between 1998 November and 2005 October, are shown graphically in 
Fig. \ref{hd187085} 
and listed in Table \ref{vel187085}. Our velocity measurements are derived 
by breaking the spectra into several hundred 2~\AA ~chunks and deriving
velocities for each chunk. The velocity error, given in the third 
column and labelled `error'
is determined from the scatter of these chunks.
This error includes the effects of photon-counting uncertainties,
residual errors in the spectrograph PSF model, and variation in
the underlying spectrum between the template and iodine epochs. All
velocities are measured relative to the zero-point defined by
the template observation. Our periodogram analysis in Fig. \ref{per187085}
reveals a single broad peak around 1000 days at about the 0.05\% false alarm
level and well above the indicated 1\% false alarm rate.

Our best fit solution yields an orbital period of 986~d, 
a semi-amplitude of stellar reflex motion (which we shorten to velocity 
amplitude) of 17~m~s$^{-1}$, and
an eccentricity of 0.47. The fit is shown in the upper part of Fig. \ref{hd187085}. 
The RMS to the Keplerian fit is 7.07~m~s$^{-1}$,
yielding a $\chi_{\nu}^2$ of 1.82. 
This solution implies a minimum ($M$~sin~$i$) mass of 0.75~\mjup, and a 
semi-major axis of 2.05~au.
However, the Keplerian curve of this eccentric
orbit includes both a peak and a sharp drop that are not sampled by our 
velocities. This raises the possibility that an orbit of lower eccentricity
might fit the data nearly as well. Orbits with lower
eccentricity deserve consideration because of the lower effective degrees of 
freedom harboured by such orbits. 
We tried other orbital solutions with fixed
eccentricities. We find a wide range of eccentricities fit the data. 
For example, a fixed eccentricity of 0.1 yields the same value of RMS
(shown in the lower part of Fig. \ref{hd187085}.  This is consistent with the numerical simulations
reported in Butler et al. (2000) which indicate the uncertainty in determining eccentricity becomes
large for low amplitude signals. The next periastron passage for 
HD187085 is 2007 June at which time the eccentricity solution can be better 
constrained.

Based on associating an object 
with others that reside near it on an H-R diagram
and finding the best-fit relationship between observed jitter and 
activity, Wright (2005)
predict a jitter of 2.8~m~s$^{-1}$ for HD~187085. 
Combining the
expected error of 3~m~s$^{-1}$ in quadrature with the expected jitter
error the expected RMS error of 4.25~m~s$^{-1}$ is rather lower than our 
plausible orbital solutions which have $\sim$7~m~s$^{-1}$ RMS. We expect these
higher RMS values arise because the stellar jitter relationship
for inactive stars is relatively poorly determined and also because
our precision for HD~187085 is limited by the 
slightly smaller equivalent widths
of its stellar lines relative to later spectral types. Thus with our
observing exposures adjusted to give S/N=200 per exposure our
precision is probably limited
to around 4 rather than 3 m~s$^{\rm -1}$. 
%We have investigated our data
%and find that the internal velocity errors correlate
%well with the number of photons per pixel.
%The velocity errors vary nearly as 1/$\sqrt{(\rm photons)}$,
%as expected from our stable stars (e.g., Butler et al. 2001).
The lack of any observed variation in
chromospheric activity between Henry et al. (1996) and Tinney et al. (2002)
or significant photometric variability within the Hipparcos dataset
gives us confidence
that the Keplerian curve arises from an exoplanet rather
than from long-period starspots or chromospherically active regions.

\section{Orbital Solution for HD~20782}

The 29 epochs of Doppler measurements for HD~20782, obtained
between 1998 August and 2005 September, are shown graphically in Fig.
\ref{hd20782} 
and listed in Table \ref{vel20782}. 
For cases of extreme eccentricity, like HD 20782,
periodograms break down.  The best-fit orbit is
found by searching over a wide-swath of period
space.  In the case of HD~20782 this is simplified
by the large velocity amplitude of the velocity variations.
                                                                          
The data are well-fit by a Keplerian curve which
yields an orbital period of $585.86\pm0.03$~d, a velocity amplitude
of $115\pm12$~m~s$^{-1}$ and
an eccentricity of $0.92\pm0.03$. The minimum (\msini) mass of the
planet is 1.77$\pm$0.22 \mjup and the semi-major axis is
$2.26\pm0.46$~au. The RMS to the Keplerian fit is 5.0~m~s$^{-1}$,
yielding a reduced ${\chi_{\nu}^2}$ of 1.01. Wright (2005)
predicts a jitter of 4.1~m~s$^{-1}$. Our errors 
come from Monte-Carlo simulations where the best fit Keplerian orbit.
is recomputed based on the noise in the data. Combining our
expected error of 3~m~s$^{-1}$ in quadrature with the expected jitter
gives an expected RMS of 5.1~m~s$^{-1}$, consistent with the 
measured value. Furthermore, the lack of any observed
chromospheric activity or photometric variations gives us confidence
that the Keplerian curve arises from an exoplanet rather
than from long-period starspots or chromospherically active regions.
The properties of the candidate
exoplanet in orbit around HD~20782 are summarised
in Table \ref{orbit}.

While we are confident about our procedures and the consistency of our solution, we find that within our errors HD~20782 shares the status of `most highly eccentric exoplanet' along with HD~80606 ($e$~=~0.9330$\pm$0.0067, Butler et al. 2006; $e$~=~0.927$\pm$0.012, Naef et al. 2001). Thus it is important that we examine the reality and our reliance on our single periastron data point (JD~-~2451000~=~1859.3054, RV~=~-197.6$\ms$).
If the time of this epoch were recorded wrong by about
3 hours, the barycentric correction would be off by about 200$\ms$ due
to diurnal rotation.  A check of the continuity of the logsheets rules this out and reveals that the periastron observations were taken under a clear sky. However, if the star were by chance observed near the Moon or in twilight the G2V reflected solar spectrum might drag the spectral line centroid off.  

At the time of observation the Moon as seen from Siding Spring was at
17$^{\rm h}$03$^{\rm m}$32$^{\rm s}$ -23$^{\rm o}$59$^{'}$57$^{"}$ (J2000 -- from the JPL Horizons System, http://ssd.jpl.nasa.gov/horizons.html). 
HD~20782 is 
at 03$^{\rm h}$20$^{\rm m}$04$^{\rm s}$ -28$^{\rm o}$51$^{'}$15$^{"}$ (J2000). Thus HD~20782 was nowhere near the Moon. In fact, the star 
is well off the ecliptic so can never get very close to the Moon.
The observation time (the mid-point of the exposure) was four 
minutes before the start of morning -18$^{\rm o}$ (astronomical) twilight.
At this time there should be no significant twilight contamination in the two ten minute exposures taken of HD~20782, indeed AAPS observations routinely continue until -12$^{\rm o}$ twilight. We are reassured of our strategy because based on our test exposures with UCLES looking at blank sky 20 minutes after -12$^{\rm o}$ twilight no sky photons are detected in a 1 min exposure (above read noise). Any solar contamination would generate a high ${\chi_{\nu}^2}$  statistic for the quality of the doppler fit. The formal internal errors for the two observations, made consecutively at the periastron epoch, were 5.1 and 5.5 m/s, compared with a median of 5.3 m/s for all observations of HD~20782.   
So, the individual spectra making up our periastron data point show no extra
scatter in velocity.                                     
                  
Despite our confidence in these two data points comprising the JD~-~2451000~=~1859.3054 epoch we also recomputed the Keplerian fit
after removing them from the data set. The resulting best-fit parameters are
period~=~599.82~d, periastron time (JD)~=~51067.098, eccentricity~=~0.732, $\omega$~=~119.6~deg, 
velocity amplitude~=~32.7~$\ms$ and planetary mass~=~0.92~$M_{\rm Jup}$.
With this new fit the period is the same as before, but the values of
eccentricity and velocity amplitude (and thus planetary mass) 
are significantly lower. So, although the qualitative sense of the orbit remains nearly the same, the epoch of low velocity data do force the velocity amplitude, eccentricity and planetary mass to be smaller than if we didn't have them. Thus, while we are confident of the extreme eccentricity of HD~20782b, we would like to 
obtain further data points near periastron.
At the last periastron (2005 March 27) we were unable to 
observe HD~20782. The constrained blocks of time
awarded to the AAPS restricts our ability to reactively schedule 
time-critical observations such as those we wish to make for HD~20782.

\section{Discussion}
The two exoplanets presented here have relatively large eccentricities 
and small velocity amplitudes.
With an eccentricity of 0.925$\pm$0.030, HD~20782 has a comparable 
eccentricity to  HD~80606 ($e$~=~0.9330$\pm$0.0067, Butler et al. 2006; 
$e$~=~0.927$\pm$0.012, Naef et al. 2001) 
which has the highest value of eccentricity reported to date. 
The companion to HD~187085 has a more modest and uncertain eccentricity. Nontheless its 
combination of large eccentricity and small velocity amplitude make it 
noteworthy. It may have one of the highest known eccentricities, alternatively, 
low eccentricity fits yield velocity amplitude solutions less than 15~$\ms$ making 
it one of only a handful of low velocity amplitude, long period exoplanet detections.
On the other hand,
if a low value of eccentricity is appropriate for the orbital solution 
of HD~187085 then velocity amplitude solutions 
of less than 15~$\ms$ will be fit placing the companion to HD~187085 
with only a handful of low velocity amplitude, long period exoplanet 
detections.

\begin{figure}
\hspace*{-0.5cm}
\includegraphics [width=80mm,angle=0]{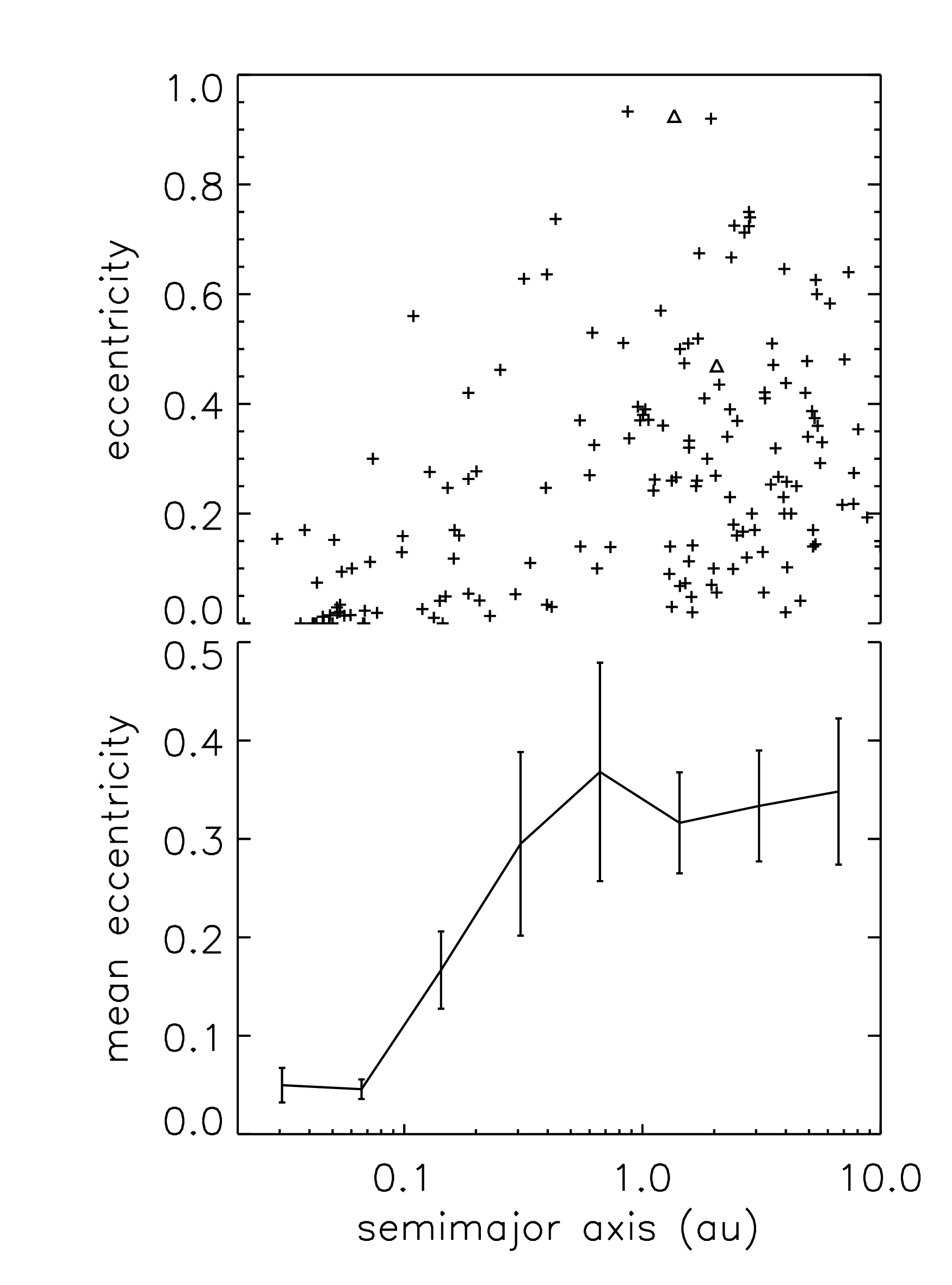}
\caption{The upper plot shows a scatter plot for eccentricities against
semimajor axis for 
exoplanets, within 200~pc, with semimajor 
axes $<$10~au and masses less than 20~$M_{\rm Jup}$ as given by table 2 of Butler et al. (2006). 
The values for the candidate exoplanets
presented here are shown as triangles. 
The lower plot shows mean eccentricities plotted against
semimajor axis for exoplanets.  Tidal circularisation dominates
the close-in planets with eccentricities reaching a plateau by around 0.2~au. 
The error bars are from $\sqrt{(number)}$ statistics
and are only indicative. The plots show 
results from many surveys and are thus drawn from an inhomogeneous sample.}
\label{aue}
\end{figure}

\begin{figure}
\hspace*{-0cm}
\includegraphics [width=80mm,angle=0]{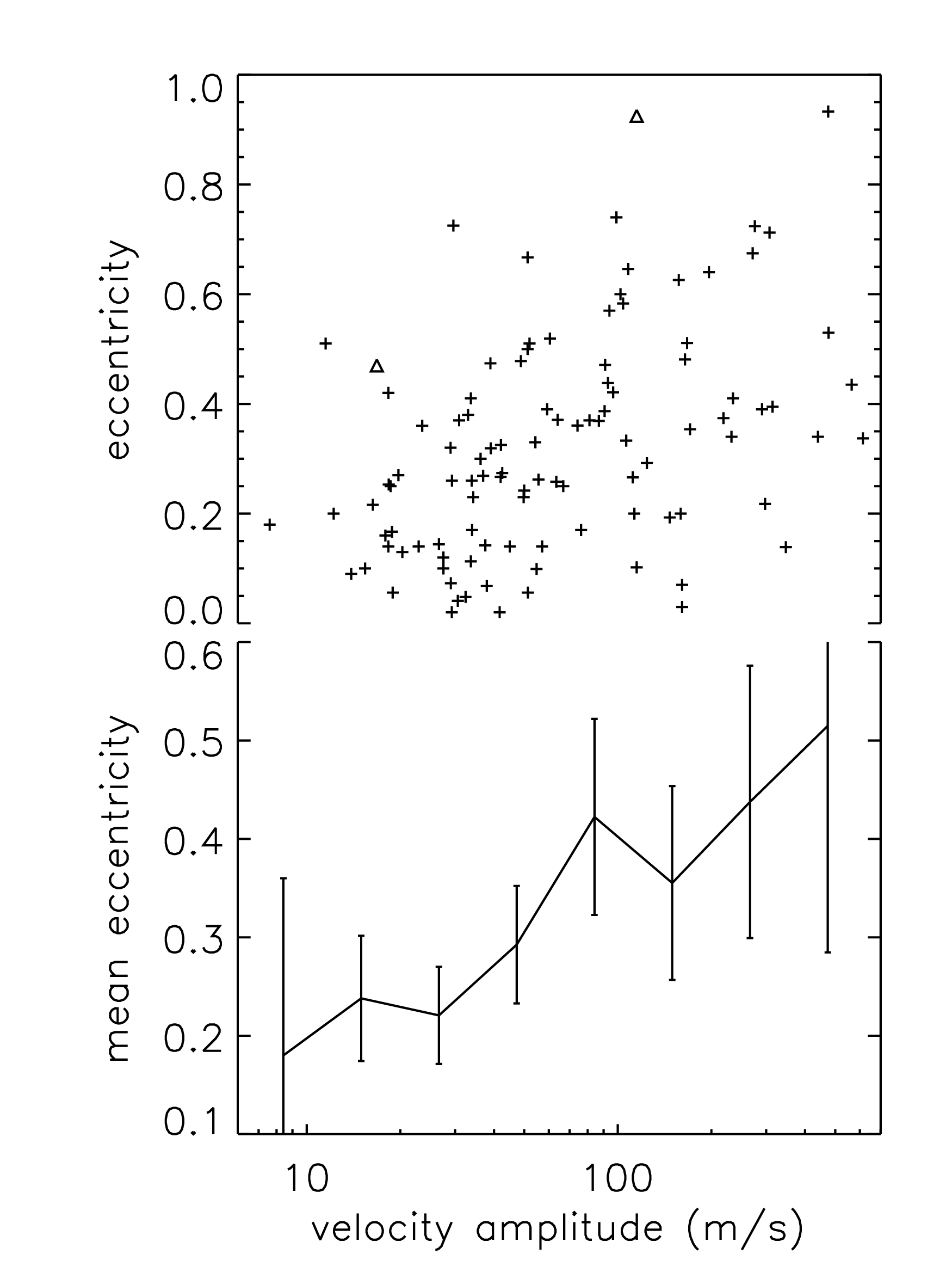}
\caption{The upper plot shows a scatter plot for eccentricities against
velocity amplitude for exoplanets, within 200~pc, with semimajor 
axes 0.5--10~au and masses less than 20~\mjup as given by table 2 of Butler et al. (2006). 
The values for the exoplanets.
The lower plot shows mean eccentricities plotted against
velocity amplitudes. The error bars are 
$\sqrt{(number)}$ statistics and are only indicative. The plots show 
results from many surveys and are thus drawn from an inhomogeneous sample.}
\label{ke}
\end{figure}

\begin{figure}
\hspace*{-0.5cm}
\includegraphics [width=80mm,angle=0]{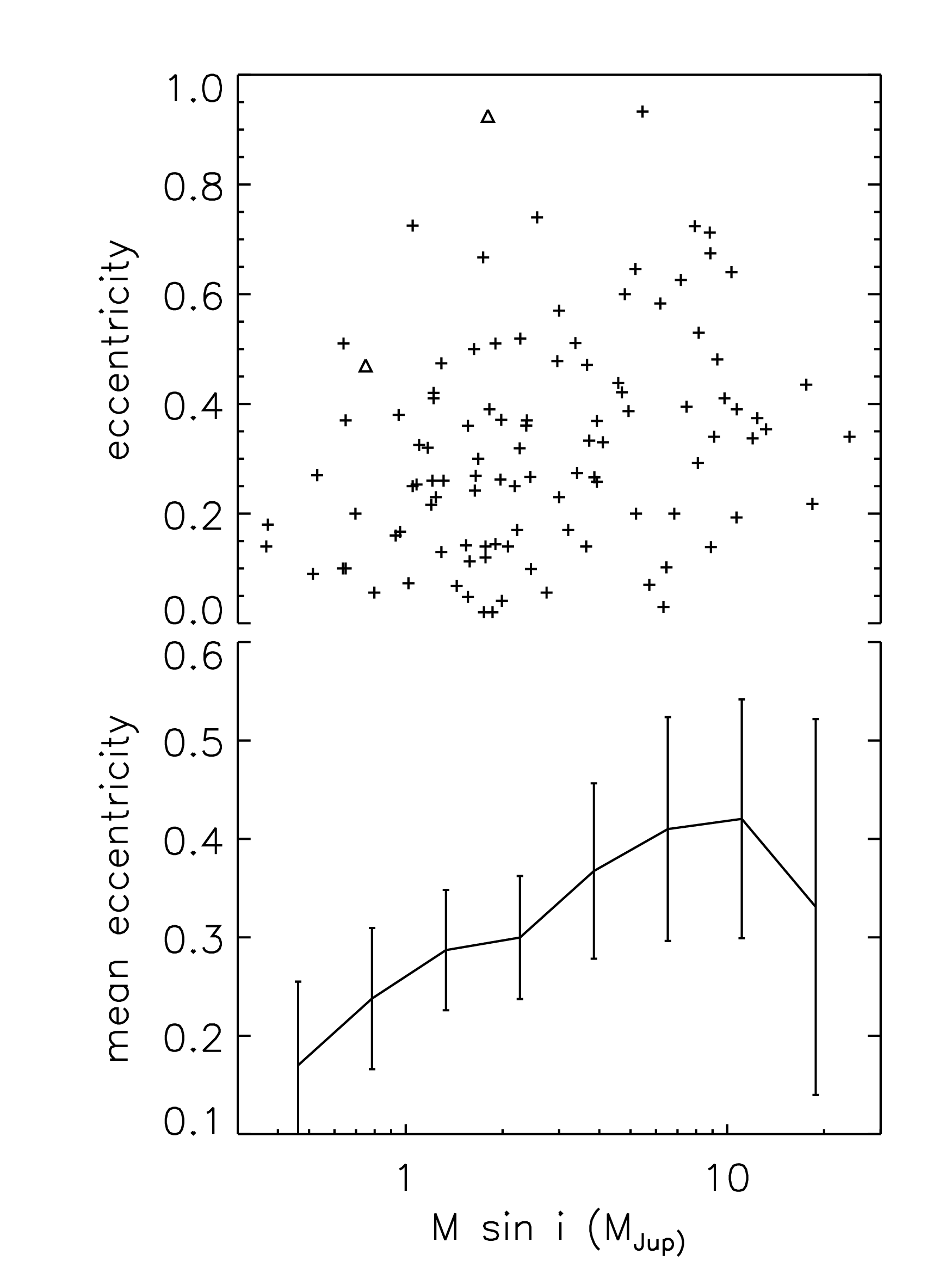}
\caption{The upper plot shows a scatter plot for eccentricities against
$M {\rm sin} i~$~$M_{\rm Jup}$ for exoplanets within 200~pc with semimajor axes 0.5--10~au and masses less than 20~$M_{\rm Jup}$ as given by table 2 of Butler et al. (2006). The values for the exoplanets
presented here are shown as triangles. The error bars are 
$\sqrt{(number)}$ statistics and are only indicative. The plots show 
results from many surveys and are thus drawn from an inhomogeneous sample.}
\label{me}
\end{figure}

The relative rarity of such high eccentricity planets can be seen 
in Fig. \ref{aue}. The plot shows a clear deficit of  
eccentric planets with small semimajor axes. For planets with periastron 
distances less than 0.1~au this deficit is expected (e.g., Rasio \& Ford 1996) and seen to be due to the tidal circularisation of 
close-in planets (e.g. Halbwachs, Mayor \& Udry 2005).
However, for periastron distances of 0.1~au the tidal circularisation timescale
is already likely to be many Gyr, for inferred $Q$ values (based on planets
having semi-major axes with less than 0.1~au, where $Q$ is the specific
dissipation function). The circularisation 
timescale, $t_{\rm circ}$, is a sensitive function of the radius of the planet, 
$R_{\rm planet}$ 
($t_{\rm circ}$ $\propto$ $R_{\rm planet}^{-5}$, Goldreich \& Soter 1966) 
and supposedly occurs due to dissipation in the planet, not the star 
(e.g., Gu et al. 2004).     
Perhaps, the larger planet radii during contraction, in
the first 10~Myr or so, helps to shorten the circularisation time. Thus the 
rising mean eccentricities versus semi-major axis apparent
in Fig. \ref{aue} from 0.1 to 0.3~au
may be explained by the decreasing effectiveness 
of the tidal circularisation mechanism.
%be a reflection of the decreasing effectiveness of the 
%tidal circularisation mechanism during the early contraction phase.
 
%In addition to a lack of highly eccentric short-period exoplanets, 
%there may also be a slight lack of highly eccentric 
%exoplanets at large semimajor axes. Plotting the mean eccentricities 
%in the lower part of
Fig. \ref{aue}, in particular the lower binned plot, indicates that
mean eccentricities rise steeply reaching a plateau beyond around 0.3~au. Thereafter,
the eccentricity of an exoplanet appears roughly constant. 
Beyond 2~au there may be a slight deficit of high eccentricity 
exoplanets apparent from the lack of objects in the top right of 
the upper part of Fig. \ref{aue}. In order to
detect highly eccentric exoplanets with long periods, 
it is necessary to have good time sampling. 
For example, the detection of the ultra-sharp periastron passages 
of eccentric orbits requires relatively large numbers of observations as 
well as fortitous sampling.
In the case of HD~20782, we only have one epoch near periastron
passage. Nonetheless, our robust observing and data reduction procedures 
and the relatively large (115 m/s) amplitude signal gives us confidence 
that we have detected a high eccentricity orbit. 
Cumming (2004) illustrates the difficulties of detecting planets
of high eccentricities with limited sampling. His work shows
that even with high signal-to-noise and representative numbers
of data points,``there remain significant selection effects
against eccentric orbits for $e~>$ 0.6."

While time sampling is crucial for adequate sampling 
of highly eccentric orbits, it is also appropriate to comment
on precision. The top left of Fig. \ref{ke} is relatively poorly 
populated. As discussed above this is likely to be due to difficulties
associated with the detection of high eccentricity orbits. In addition,
there maybe a relatively smaller population of high eccentricity exoplanets.. 
However, at more modest eccentrcities,
there is an apparent deficit of low velocity amplitude exoplanets.
It is interesting to look at the objects that define the upper envelope
of the top left of the upper part of Fig. \ref{ke}.
These objects represent
the highest eccentricity objects that have been found for a given 
velocity amplitude. In addition to the companion to HD~20782 reported here, 
the other two objects are 
HD~45350~b (Marcy et al. 2005a) and HD~11964~b (Butler et al. 2006), both
recently discovered using Keck data. 
Both the AAPS and Keck planet search operate with  
demonstrated long-term precisions of around 3 m/s (e.g. fig. 1, Tinney et al.
2005 and fig, 2, Vogt et al. 2005). 
Such long-term precisions may now be rivalled over shorter baselines by the  
HARPS (e.g., Lovis et al. 2005) and Hobby-Eberly (e.g., McArthur et al. 2004) 
planet search projects though 
many of the objects plotted in Fig. \ref{ke} were discovered 
by searches with lesser long term precision.
While the importance of precision is 
widely recognised its specific importance for the detection of high 
eccentricities has been anticipated 
in the simulations of Cumming (2004) and Halbwachs et al. (2005). 
Cumming's fig. 4 shows that 
the detectability of exoplanets with $e>0.7$ is a function of 
precision. Thus, simulations and the exoplanets detected so far
serve to suggest that radial velocities with good long-term 
precision are important to robustly sample the parameter space occupied by
highly eccentric planets.

The observational difficulties inherent in the detection of
low velocity amplitude -- high eccentricity planets predicted by Cumming
and suggested in the upper part of Fig. \ref{ke} appear to be borne out 
by the binned (lower part) of the figure.
These binned data suggest that the velocity amplitude of exoplanets increases 
with mean eccentricity in Fig. \ref{ke}, albeit with low confidence since
it is almost possible to draw a flat line through all the error bars. 
%, rather than the insensitivity
%expected for no relationship between velocity and eccentricity.
%Nonetheless, since it is still 
%possible to draw a flat line through all the error bars, this trend does
%not yet have any strong statistical significance. 
While we have attempted to remove the effects of 
circularisation processes from Fig. \ref{ke} 
(by removing all planets with semi-major axes less than 0.5~au), it is 
plausible that the suggested trend may have a physical origin.
In particular, it is notable that beyond the tidal circularisation limit 
(usually assumed to be 0.1~au) spectroscopic binaries have significantly 
higher eccentricities than exoplanets (Halbwachs et al. 2005) and 
that high mass planets (also having high velocity amplitudes) 
have systematically higher eccentricities than low mass planets,
 e.g., 70 Vir (Marcy \& Butler 1996) and fig. 5, Marcy et al. (2005b).
Fig. \ref{me} shows this correlation of eccentricity with mass for the exoplanet sample of Butler et al. (2006). For large
velocity amplitudes and large mass planets this cannot be a selection 
effect nor can it be caused by errors because the most massive 
planets induce the largest Keplerian amplitudes, allowing precise determination
of eccentricity.
%However, in the absence of detailed simulations within the observed 
%data sets, the observed trend appears to support the hypothesis 
%that for a given velocity amplitude there is an observational bias against 
%the detection of highly eccentric exoplanets. 
To quantify the reality of the suggested trends, 
to assess observational bias at high eccentricities 
and to recover the underlying 
eccentricity distribution of exoplanets analysis of the radial 
velocity data for each different survey is required. Such work 
has been started by Cumming et al. (1999, 2003) and independently by the Anglo-Australian Planet Search.

By the standards of our Solar System, the majority of exoplanets 
beyond 1~au can be seen from Fig. \ref{aue} to be in eccentric orbits. 
This contrast with the Solar System may be at least partly explained by 
as yet unseen long-period 
companions. Indeed the fits for a few of the long-period eccentric planets 
do already include longer-period trends though these are generally rather 
modest and sometimes uncertain, e.g., our `best fit' solution for 
HD~187085 includes an additional slope of 1.3$\pm$1.0~\ms. 
So while individual cases require careful consideration, 
the lack of substantial trends in the current 
sample of high eccentricty
exoplanets (e.g., Butler et al. 2006) means that the majority of the 
observed eccentricities need to be explained 
by high formation eccentricties or by appropriate 
eccentricity pumping mechanisms. 
Since the derived masses of exoplanets are directly proportional to 
measured velocity amplitude it is important to account for the likely
observational eccentricity bias when investigating mass dependencies
of exoplanets and in particular when trying to understand the relative
importance of different eccentricity pumping 
mechanisms (e.g., Takeda \& Rasio 2005).
%or mass dependence (e.g. Mazeh, Mayor \& Latham 1997).
We are encouraged to proceed in earnest with our project to 
examine the detectability of Anglo-Australian Planet Search exoplanets 
of different 
eccentricities (and other parameters), given our errors and time sampling. 
This should improve our search and will put us 
in a better position to better understand the underlying distribution
of eccentricity with mass and semimajor axis for exoplanets.

\section*{Acknowledgments}
We thank the referee, Gordon Walker for his valuable comments.
We are very grateful 
for the support of the Director of the AAO, Dr Matthew Colless,
and the superb technical support which has been received throughout
the programme from AAT staff -- in particular John Collins, Shaun James, 
Steve Lee, Rob Patterson, Jonathan Pogson and Darren Stafford. 
We gratefully
acknowledge the UK and Australian government support of the
Anglo-Australian Telescope through their PPARC and DETYA funding
(HRAJ, AJP, CGT); NASA grant NAG5-8299 \& NSF grant
AST95-20443 (GWM); NSF grant AST-9988087 (RPB); and
Sun Microsystems.  This research has made use of the SIMBAD
database, operated at CDS, Strasbourg, France.

\clearpage
\end{document}